\documentclass{elsart}
\usepackage{amssymb,graphicx}
\usepackage{amsmath}
\usepackage{psfrag,epsfig}

\begin{document}

\begin{frontmatter}

\title{Proton incorporations and superconductivity in a cobalt oxyhydrate}

\author[ZJU_P]{Guanghan Cao\corauthref{cor1}}\ead{ghcao@zju.edu.cn},
\author[ZJU_T]{Xiaoming Tang},
\author[ZJU_P]{Yi Xu},
\author[ZJU_P]{Ming Zhong},
\author[ZJU_P]{Xuezhi Chen},
\author[ZJU_T]{Chunmu Feng},
and \author[ZJU_P]{Zhu'an Xu} \corauth[cor1] {Corresponding author}
\address[ZJU_P]{Department of Physics, Zhejiang University,
Hangzhou, Zhejiang 310027, People's Republic of China}
\address[ZJU_T]{Test and Analysis Center, Zhejiang University,
Hangzhou, Zhejiang 310027, People's Republic of China}

\date{\today}

\begin{abstract}
We report the evidence of proton incorporations in a newly-discovered cobalt oxyhydrate superconductor. During
the hydration process for Na$_{0.32}$CoO$_{2}$ by the direct reaction with water liquid, it was shown that
substantial NaOH was gradually liberated, indicating that H$^{+}$ is incorporated into the hydrated compound.
Combined with the thermogravimetric analysis, the chemical composition of the typical sample is
Na$_{0.22}$H$_{0.1}$CoO$_{2}\cdot 0.85$H$_{2}$O, which shows bulk superconductivity at 4.4 K.
\end{abstract}


\end{frontmatter}


Recently, Takada et al.~\cite{Takada} discovered the first cobalt oxide superconductor denoted as
Na$_{x}$CoO$_{2}\cdot y$H$_{2}$O ($x\approx0.35, y\approx1.3$). Though the superconducting transition
temperature $T_{c}$ ($\sim$ 5 K) is not striking, the underlying physics is so attractive that many research
groups have been following this
topic~\cite{Baskaran,Kumar,Wang,Tanaka,Ytanaka,Foo,Schaak,Cmaidalka,Sakurai,Milne,Chen,Cao,Park,Lynn,Jorgensen,Jin}.
Various audacious ideas and predictions have been proposed by the
theorists~\cite{Baskaran,Kumar,Wang,Tanaka,Ytanaka}, waiting for the experimental verifications. On the other
hand, experimental studies were found to be very difficult primarily due to the extreme chemical instability of
the compound~\cite{Foo}. Inconsistent experimental results often appear in the literatures. Therefore, it is
essential to carefully characterize the sample before the physical, structural, and spectral properties are
measured. Up to present, the chemical characterizations of the sample mainly focus on the sodium and water
content~\cite{Schaak,Cmaidalka,Sakurai,Milne,Chen}. In this paper, we report the evidence of proton
incorporations in the new superconducting cobalt oxyhydrate.

Similar to the previous reports~\cite{Takada,Foo}, samples were prepared in the following steps. First,
Na$_{0.7}$CoO$_{2}$ polycrystals were synthesized by three rounds of fast solid-state reaction at 1083 K in
flowing oxygen with two intermediate regrindings, using Na$_{2}$CO$_{3}$ (99.9\%) and Co$_{3}$O$_{4}$ (99.99\%)
as the starting materials. Although the original composition is Na$_{0.74}$CoO$_{2}$, the actual composition was
determined to be Na$_{0.7}$CoO$_{2}$ by the atomic absorption spectroscopy (AAS) and the Na$^{+}$ ion-selective
electrode (ISE) techniques. The loss of sodium is primarily due to the volatilization of Na$_{2}$O during the
solid-state reaction. Fig. 1(a) shows the powder x-ray diffraction (XRD) pattern for Na$_{0.7}$CoO$_{2}$, as
measured with Cu K$\alpha$ radiation. All the diffraction peaks can be well indexed using a hexagonal cell with
$a$=2.831 \AA\ and $c$=10.918 \AA\ .

\begin{figure}
\includegraphics[width=12cm]{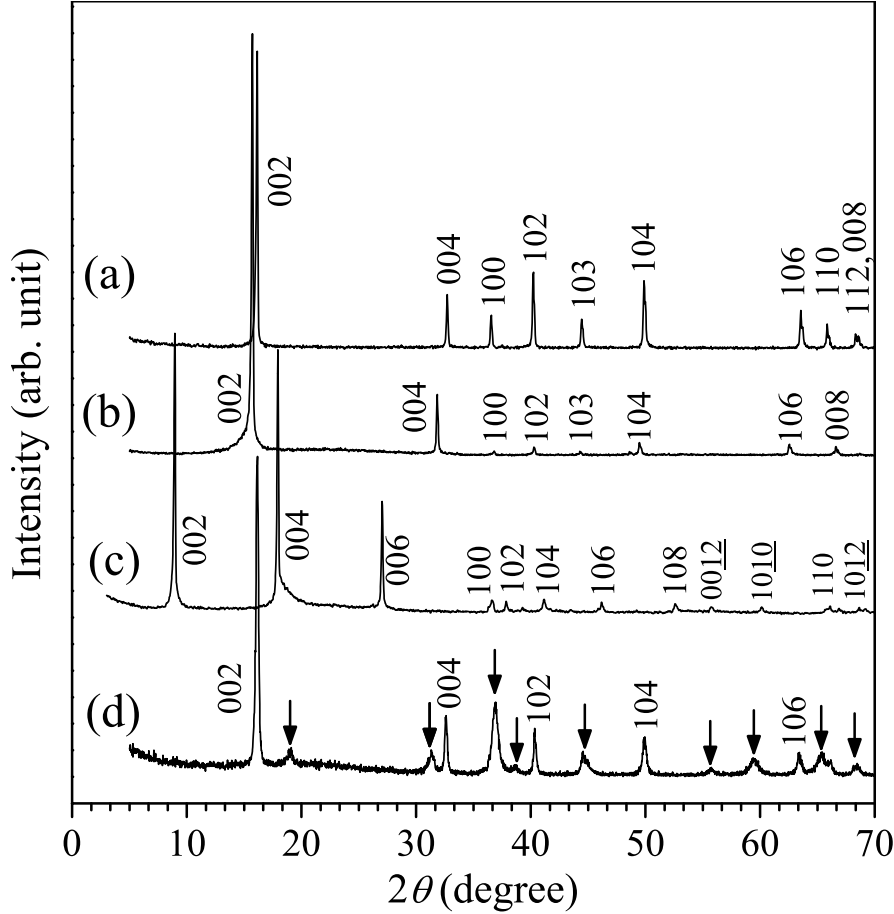}
\caption{X-ray diffraction patterns of (a) Na$_{0.7}$CoO$_{2}$, (b) Na$_{0.32}$CoO$_{2}$, (c) the
water-liquid-hydrated (WLH) sample, and (d) as-fired WLH sample at 673 K in air for 0.5 hours. The diffraction
indices for the peaks are labelled. In the pattern (d), the peaks marked by arrows indicate that Co$_{3}$O$_{4}$
is separated out.}
\end{figure}

In the second step, Na$_{0.7}$CoO$_{2}$ was oxidized by the excessive bromine dissolved in acetonitrile. This
process resulted in the deintercalation of the sodium, producing the thermodynamically metastable hexagonal
phase Na$_{x}$CoO$_{2}$ ($0.25<x<0.7$). The value of $x$ depends on the equilibrium concentrations of Br$_{2}$,
Br$^{-}$, and Na$^{+}$, provided the reaction time is long enough (over 48 hours for polycrystalline samples).
In the case of the sample preparation in the present study, 2.140 g Na$_{0.7}$CoO$_{2}$ reacted with 3.70 g
Br$_{2}$ dissolved in 10.0 ml acetonitrile  in a closed container at 310 K for 48 hours. After the reaction
completed, the solid product was washed several times with acetonitrile and then dried in vacuum. The mass of
the solid Na$_{x}$CoO$_{2}$ became 1.965 g. From the loss of mass, one can estimate that the $x$=0.32, which is
in very good agreement with our ISE measurement result. Fig. 1(b) shows the XRD pattern for the sample
Na$_{0.32}$CoO$_{2}$. It has the same crystal structure with that of Na$_{0.7}$CoO$_{2}$, but the unit cell
becomes elongated: $a$=2.811 \AA\ and $c$=11.211 \AA\ . The shrinkage of $a$-axis is ascribed to the increase of
the oxidation state of cobalt, and the stretch of $c$-axis is due to the relatively weak Coulomb attractions
along the $c$-axis when Na$^{+}$ is partially deintercalated.

The intermediate compound Na$_{0.32}$CoO$_{2}$ easily absorbs water. So, most literatures employed the reaction
with water vapor for the hydration. However, deliquescence often happens, making it difficult to distinguish
between crystal water and free water. Now that free water inclusion is inevitable, one can prepare the hydrated
compound by the direct reaction with water liquid or solution, and it was proved to be
successful~\cite{Cao,Park}. In this study, we employed both routes to synthesize the hydrated compound in order
to make a comparison. Partial Na$_{0.32}$CoO$_{2}$ sample reacted with water vapor at room temperature for one
week, obtaining the water-vapor-hydrated (WVH) compound. Another part of Na$_{0.32}$CoO$_{2}$ was soaked with
water liquid in a closed container at room temperature for one week. This product is hereafter called
water-liquid-hydrated (WLH) sample. XRD patterns of the two hydrated products are quite similar. By the
least-squared fitting, the cell parameters were calculated as $a$=2.823 \AA\ and $c$=19.61 \AA\ for the WVH
compound, consistent with the previous report~\cite{Takada}. However, the cell parameters for the WLH product
are $a$=2.824 \AA\ and $c$=19.75 \AA\ . The obvious difference in the value of $c$-axis implies some minute
change in the crystal structure. We will discuss this issue later. Fig. 1(c) shows the XRD pattern of the WLH
sample, which indicates that it is a hexagonal single phase.

The thermal instability of samples was investigated by using a thermal analyzer which can simultaneously measure
the temperature dependence of weight (thermogravimetric analysis, TGA) and the temperature difference between
the sample and the reference (differential thermal analysis, DTA). The experiments were performed under ambient
condition ($T$=298 K, humidity: $\sim50 \%$) using the sweep rate of 20 K/min. Fig. 2 shows the TGA result for
the different samples. The first curve for Na$_{0.7}$CoO$_{2}$ shows an abrupt loss of weight at 1300 K,
accompanied with a big endothermal peak in the DTA curve (not shown here). The measured sample was found to have
molten after cooling down. Therefore, the decomposition can be expressed as
\begin{equation}
\mathrm{Na}_{0.7}\mathrm{CoO}_{2}\longrightarrow\mathrm{Liquid}
 (0.35\mathrm{Na}_{2}\mathrm{O}+\mathrm{Co}\mathrm{O})+0.325\mathrm{O}_{2}\uparrow.
\end{equation}

\begin{figure}
\includegraphics[width=12cm]{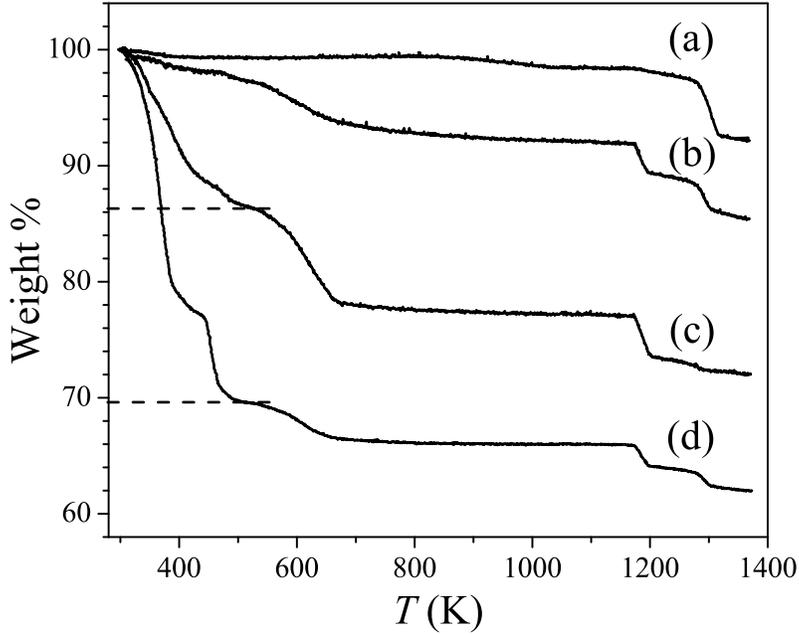}
\caption{Thermogravimetric curves of (a) Na$_{0.7}$CoO$_{2}$, (b) Na$_{0.32}$CoO$_{2}$, (c) the
water-liquid-hydrated compound, and (d) the water-vapor-hydrated compound.}
\end{figure}

In the second curve for Na$_{0.32}$CoO$_{2}$, there are weight-losses at 605 K, 1184 K, and 1291 K,
respectively. XRD measurement for the sample annealed at 673 K in air shows that Co$_{3}$O$_{4}$ is separated
out, like the case in Fig. 1(d). So, the loss of weight at $\sim$ 605 K corresponds to the following equation,
\begin{equation}
\mathrm{Na}_{0.32}\mathrm{CoO}_{2}\longrightarrow0.457\mathrm{Na}_{0.7}\mathrm{CoO}_{2}+0.181\mathrm{Co}_{3}\mathrm{O}_{4}+0.181\mathrm{O}_{2}\uparrow.
\end{equation}
Since Co$_{3}$O$_{4}$ decomposes at about 1180 K in air, the weight-loss at 1184 K is ascribed to the
decomposition of Co$_{3}$O$_{4}$,
\begin{equation}
\mathrm{Co}_{3}\mathrm{O}_{4}\longrightarrow3\mathrm{Co}\mathrm{O}+0.5\mathrm{O}_{2}\uparrow.
\end{equation}
It is noted that the theoretical weight-losses based on Eqs. (1), (2) and (3) are basically consistent with the
experimental results.

As for the hydrated samples, heavy weight-loss was observed below 518 K, which is due to the loss of water.
Around 600 K, the weight-loss is associated with the decomposition of the metastable Na$_{x}$CoO$_{2}$, as
indicated by Fig. 1(d). The other two weight-losses at higher temperatures can be described by the Eqs. (1) and
(3), respectively. Since the sodium content was determined to be 0.22 for the WLH sample (see below), the water
content for this compound is thus estimated to be 0.85. This value is obviously lower than the common value
$\sim$ 1.3~\cite{Takada,Foo,Cmaidalka,Chen}. On the contrary, the water content for the WVH compound is
determined as 2.3, which is remarkably higher than that of previous reports. We suspect that free water had been
absorbed in the WVH sample. It is noted that the ratio of weight-loss at 1184 K and 1291 K is quite different
for the two hydrated samples. This is because that the sodium content in the WLH compound is lower than that in
the WVH sample.

The sodium content in the cobaltates was measured by AAS and Na$^{+}$ ISE techniques, respectively. In the AAS
method, the sample was dissolved in 2 mol/L HNO$_{3}$ solution and then diluted into appropriate concentrations
for the measurement. Blank data (parallel experiment result with no sample dissolved) were collected and then
deducted because the Na concentration in the HNO$_{3}$ solution is generally not neglectable. In the ISE
measurement, the sample was dissolved in acid solution, and then the solution was neutralized by
(CH$_{3}$)$_{2}$CHNH$_{2}$ to eliminate the disturbance of H$_{3}$O$^{+}$. Both measurements show that the
sodium content for the WLH compound is 0.22(1), which is 1/3 smaller than that of the WVH one.

We also used the ISE technique to determine the Na$^{+}$ concentration dynamically. Fig. 3 shows the change of
Na$^{+}$ concentration against the time of the hydration reaction at 300 K. It can be seen that the liberated
Na$^{+}$ increases with increasing the reaction time. During the first 10 hours of the hydration, Na$^{+}$
concentration increases rapidly. Then, it increases gradually until the saturation at 0.01 mol/L when $t$=120
hours. It was also noted that the pH value increases from 9.0 at the beginning of the hydration to 12.0 when the
reaction completed. That is to say, NaOH is gradually liberated during the hydration reaction. Considered that
the hydration is not a redox reaction, therefore, H$^{+}$ has to be incorporated in the hydrated product. By
using the above TGA result, the WLH compound can be expressed as Na$_{0.22}$H$_{0.1}$CoO$_{2}\cdot
0.85$H$_{2}$O, and the hydration can be described as
\begin{equation}
\mathrm{Na}_{0.32}\mathrm{CoO}_{2}+0.95\mathrm{H}_{2}\mathrm{O}\longrightarrow\mathrm{Na}_{0.22}\mathrm{H}_{0.1}\mathrm{CoO}_{2}\cdot
0.85\mathrm{H}_{2}\mathrm{O}+0.1\mathrm{Na}^{+}+0.1\mathrm{OH}^{-}.
\end{equation}
Apparently, the equation above consists of the ion-exchange of Na$^{+}$ and H$^{+}$. So, it is not strange that
the content of the incorporated proton was found to alter with the change of the concentration of NaOH in the
reactor. When the NaOH was removed, the content of the incorporated proton could increase up to 0.15. On the
contrary, in the case of high NaOH concentration, the amount of the incorporated H$^{+}$ will be reduced. Note
that free water may exist in the WVH sample, we speculate that small amount of proton might be also incorporated
in the Na$_{0.32}$CoO$_{2}\cdot 2.3$H$_{2}$O phase.

\begin{figure}
\includegraphics[width=12cm]{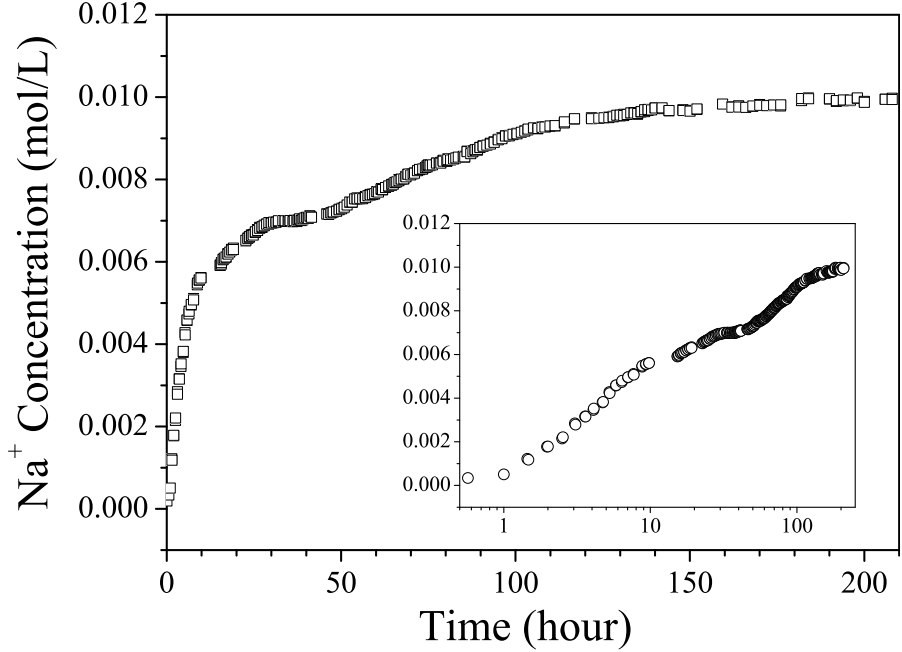}
\caption{Na$^{+}$ concentration as a function of the hydration reaction time at 300 K. Note that the inset uses
the logarithm scale for the horizontal axis.}
\end{figure}

Now, let us discuss the possible crystal structure of Na$_{0.22}$H$_{0.1}$CoO$_{2}\cdot 0.85$H$_{2}$O. As we
know, the anhydrus parent compound Na$_{0.32}$CoO$_{2}$ consists of triangular CoO$_{2}$ layers in which cobalt
is octahedrally coordinated. Partially-occupied Na$^{+}$ is sandwiched by the CoO$_{2}$ layers. Water can be
intercalated between CoO$_{2}$ layers and Na$^{+}$ layers when the sodium is deintercalated to some extent. By
using neutron diffractions, detailed structural models were established~\cite{Lynn,Jorgensen}. Impressively,
Jorgensen et. al.~\cite{Jorgensen} proposed that the position of Na$^{+}$ was shifted in such a way to
accommodate the tetrahedral coordinations by water molecules. In this model, the ideal Na to H$_{2}$O ratio is
1:4, which also satisfies the chemical formula Na$_{0.22}$H$_{0.1}$CoO$_{2}\cdot 0.85$H$_{2}$O. So, we think
that similar structure is probable for the present WLH compound. Another problem concerns about the position of
the incorporated H$^{+}$. From chemical bonding point of view, H$^{+}$ may bond with the oxygen in H$_{2}$O to
form H$_{3}$O$^{+}$, or, bond with the oxygen in CoO$_{2}$ layers to form the OH groups. We have measured the IR
spectra, but no definite conclusion can be drawn about the bonding of the H$^{+}$.

The longer $c$-axis of Na$_{0.22}$H$_{0.1}$CoO$_{2}\cdot 0.85$H$_{2}$O compared with that of
Na$_{0.32}$CoO$_{2}\cdot 2.3$H$_{2}$O can be explained in terms of Coulomb attractions between the Na$^{+}$
layers and CoO$_{2}$ layers. One would expect that relatively weak Coulomb attractions along the $c$-axis for
Na$_{0.22}$H$_{0.1}$CoO$_{2}\cdot 0.85$H$_{2}$O due to the relatively less electric charge in the Na$^{+}$ and
CoO$_{2}$ layers. So, though the water content is not high in Na$_{0.22}$H$_{0.1}$CoO$_{2}\cdot 0.85$H$_{2}$O,
the $c$-axis parameter can be even larger than that of Na$_{0.32}$CoO$_{2}\cdot 2.3$H$_{2}$O.

The superconducting transition was investigated on a Quantum Design PPMS system. As shown in Fig. 4, the
as-prepared WVH and WLH samples shows diamagnetic transition at 4.5 K and 4.4 K, respectively. The sharp
transitions indicate that both samples have bulk superconductivity, though the diamagnetic signal for the sample
WVH is stronger. This implies that the proton incorporation does not influence the superconductivity so much.

\begin{figure}
\includegraphics[width=12cm]{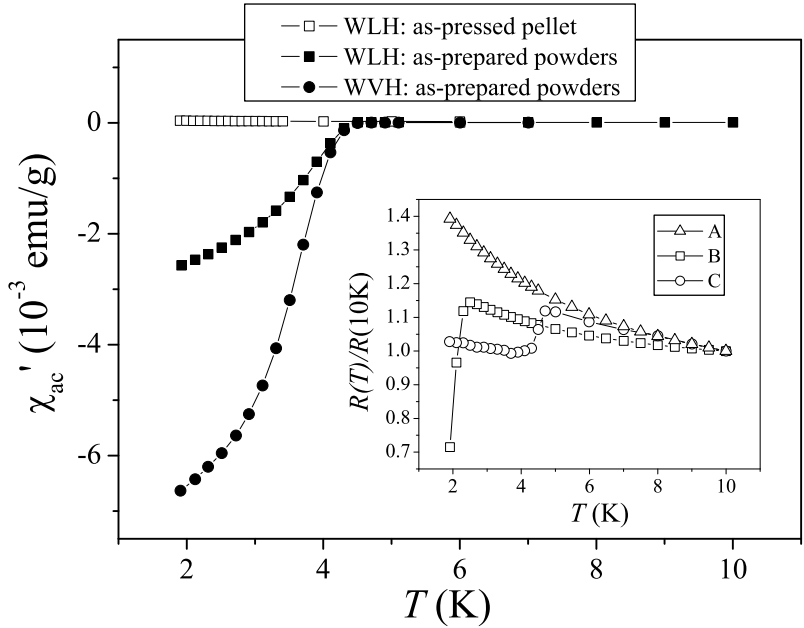}
\caption{Superconducting transitions measured by ac susceptibility and dc resistance. In the inset, data A was
collected in a fresh pellet pressed with the pressure of 5000 kg/cm$^{2}$. Data B and C were measured after the
pellet was placed in humid conditions at room temperature for one day and two weeks, respectively.}
\end{figure}

It was found that the superconductivity in Na$_{0.22}$H$_{0.1}$CoO$_{2}\cdot 0.85$H$_{2}$O is very sensitive to
the details of the measurement operations. While the as-prepared WLH powders show bulk superconductivity at 4.4
K, the as-pressed pellets (with the pressure of 5000 kg/cm$^{2}$) shows no superconducting transition above 1.9
K. The resistance measurement shown in the inset of Fig. 4 confirms the result in an opposite way. The fresh
pellet shows no superconducting transition above 1.9 K, however, after the very same sample was placed in humid
environment at room temperature over a period of time, superconductivity was recovered. Moreover, $T_{c}$
increases with increasing the placement time. Therefore, it should be careful to draw any conclusions on the
relationship between $T_{c}$ and the factors, such as water and sodium content. We checked the XRD patterns for
the as-pressed and the as-placed pellets. Both data indicates single phase of $c$=19.7 \AA\ , however, the
as-pressed sample shows very broad diffraction peaks, and the broad peaks change back into sharp ones for the
as-placed sample. This observation suggests that the crystallinity, affected by water, be important for the
superconductivity.

In summary, the water-liquid-hydrated cobaltate sample, which shows bulk superconductivity, was carefully
characterized by the XRD, ISE, TGA measurements. Evidence of proton incorporations has been given, though the
position and the chemical bonding of the hydrogen ions are not clear. It was observed that the superconductivity
is extremely sensitive to temperature, pressure and humidity. Crystallinity seems to play a role in the
appearance of superconductivity.

\textbf{Acknowledgements}

This work was supported by NSFC under Grant No. 10104012 and 10225417. Partial support under Project No.
NKBRSF-G1999064602 is also acknowledged.

\end{document}